\documentclass[useAMS,usenatbib,usegraphicx]{mn2e}

\title[The accretion disc in MG 0414+0534]{A microlensing study of the accretion disc in the quasar MG 0414+0534\thanks{This paper includes data gathered with the 6.5 meter Magellan Telescopes located at Las Campanas Observatory, Chile.}}
\author[N. F. Bate, D. J. E. Floyd, R. L. Webster and J. S. B. Wyithe]{N. F. Bate$^{1}$\thanks{E-mail: nbate@physics.unimelb.edu.au (NFB); dfloyd@lco.cl (DJEF), rwebster@physics.unimelb.edu.au (RLW); swyithe@physics.unimelb.edu.au (JSBW)}, D. J. E. Floyd$^{1,2}$$\dagger$, R. L. Webster$^{1}$$\dagger$ and J. S. B. Wyithe$^{1}$$\dagger$\\
$^{1}$School of Physics, The University of Melbourne, Parkville, Vic, 3010, Australia\\
$^{2}$OCIW, Las Campanas Observatory, Casilla 601, Colina El Pino, La Serena, Chile}
\begin{document}

\date{Accepted 2008 September 30. Received 2008 September 21; in original form 2008 July 23}

\pagerange{\pageref{firstpage}--\pageref{lastpage}} \pubyear{2008}

\maketitle

\label{firstpage}

\begin{abstract}
Observations of gravitational microlensing in multiply imaged quasars currently provide the only direct probe of quasar emission region structure on sub-microarcsecond scales. Analyses of microlensing variability are observationally expensive, requiring long-term monitoring of lensed systems. Here we demonstrate a technique for constraining the size of the quasar continuum emission region as a function of wavelength using single-epoch multi-band imaging. We have obtained images of the lensed quasar MG 0414+0534 in five wavelength bands using the Magellan 6.5-metre Baade telescope at Las Campanas Observatory, Chile. These data, in combination with two existing epochs of \textit{Hubble Space Telescope} data, are used to model the size of the continuum emission region $\sigma$ as a power-law in wavelength, $\sigma\propto\lambda^\nu$. We place an upper limit on the Gaussian width of the $r^\prime$-band emission region of $1.80 \times 10^{16} h_{70}^{-1/2}(\langle M \rangle/\rmn{M}_{\odot})^{1/2}$cm, and constrain the power-law index to $0.77\leq\nu\leq2.67$ (95 per cent confidence range). These results can be used to constrain models of quasar accretion discs. As a example, we find that the accretion disc in MG 0414+0534 is statistically consistent with a Shakura-Sunyaev thin disc model.
\end{abstract}


\section{Introduction}
Models of quasar accretion are currently poorly constrained by observations (\citealt{b07}). Accretion discs are much too small for direct imaging, and so other observational methods are required to probe their structure. Gravitational microlensing of multiply imaged quasars is one such technique, uniquely able to probe accretion disc structure on scales $\sim10^{16}$cm.

The quasar central engine is thought to consist of an accretion disc surrounding a supermassive black hole. It is a general feature of accretion disc models (e.g. \citealt{ss73}; \citealt{hh97}) that longer wavelength radiation is emitted from larger regions in the disc. Accretion discs are thus a perfect target for gravitational microlensing analyses, as the amplitude of fluctuations produced by gravitational microlensing depends strongly on the size of the emission region in the source. In particular, microlensing models predict larger fluctuations at shorter wavelengths (e.g. \citealt{rb91}; \citealt*{jwp92}; \citealt{ak99}; \citealt{yetal99}).

Considerable effort has been invested in microlensing analyses to constrain the sizes of quasar emission regions (e.g. \citealt*{wps90b}; \citealt{rb91}; \citealt*{wms95}; \citealt{wwtm00}; \citealt*{wow05}; \citealt{metal06}; \citealt{cetal08}). Recent efforts have focussed on observations in (typically) two wavelength bands, in an effort to better constrain models of quasar accretion. These analyses frequently make use of the method presented in \citet{k04}.

\citet{pooley07} analysed ten quadruply imaged quasars (including MG 0414+0534, the quasar studied in this paper), and argued qualitatively that optical emission regions must be larger than expected from basic thin-disc models by factors of $\sim3-30$. This result was supported by \citet{metal07}, in which accretion disc sizes were determined for ten lensed quasars, and \citet{metal08}, which focussed on X-ray and optical monitoring of PG 1115+080. \citet{timo08} used \textit{g'}- and \textit{r'}-band monitoring of Q2237+0305 to obtain a spectral slope consistent with the standard thin-disc models. \citet*{poin08} also found an accretion disc scale and slope consistent with standard thin disc models, using 13 years of photometric data of the two-image lens HE 1104-1805.

The quantitative analyses discussed above rely upon monitoring of microlensing-induced brightness fluctuations over an extended time period. Constraints on the size of the emission region in the source are obtained by producing a large number of simulated microlensing light curves, and searching for fits to the observational data. These constraints are therefore dependent on the transverse velocity, which can be estimated from the light curve (\citealt{wwt00a}; \citealt{k04}). In \citet*[][hereafter BWW07]{bww07} we demonstrated an alternative technique for placing constraints on the size of quasar emission regions using single observations of lensed quasars displaying a flux ratio anomaly. 

Several gravitationally lensed quasars are observed to contain a pair of images straddling a critical curve. Basic lensing theory predicts that these images should have the same magnification. However in several cases, one image is observed to be significantly demagnified relative to the other (\citealt{cr79}; \citealt{bn86}; \citealt{wms95}). It has been argued that microlensing by a combination of smooth and clumpy matter in the lensing galaxy is the most likely explanation for this discrepancy \citep{sw02}. 

By exploring microlensing simulations with varying source size and smooth matter content in the lens, we have previously obtained a 95 per cent upper limit on the size of the \textit{I}-band emission region in the anomalous lensed quasar MG 0414+0534 of $2.62 \times 10^{16} h^{-1/2}_{70} (M/\rmn{M}_{\odot})^{1/2}$ cm (BWW07). As with all lengths measured from microlensing simulations, this size is normalised to the Einstein Radius, which is scaled by the square root of the microlens mass $M$. 

In this paper, we extend our analysis across multiple filters. Images of MG 0414+0534 in the $r^\prime$, $i^\prime$, $z^\prime$, $J$ and $H$ filters were obtained on 2007 November 3, using the Magellan 6.5-metre Baade telescope. We also make use of \textit{Hubble Space Telescope} (\textit{HST}) data in the NICMOS F110W and F205W filters, taken on 1994 August 14\footnote{http://cfa-www.harvard.edu/castles/}, and the WFPC2 F675W and F814W filters, taken on 1994 November 8 \citep*{fls97}. These data allow us to obtain emission region size constraints in each filter or, equivalently, a power-law fit to the radius of the continuum emission region in the quasar as a function of emitted wavelength.

The observations are described in Section 2. In Section 3 we summarise our simulation method, with particular emphasis on how it varies from that presented in BWW07. In Section 4 we present the resulting power-law fit to accretion disc radius as a function of emitted wavelength, and discuss our conclusions.

Throughout this paper we use a cosmology with $H_0=70\rmn{kms^{-1}Mpc^{-1}}$, $\Omega_m=0.3$ and $\Omega_{\Lambda}=0.7$.

\section{Observations}
\begin{figure*}
  \includegraphics[width=170mm]{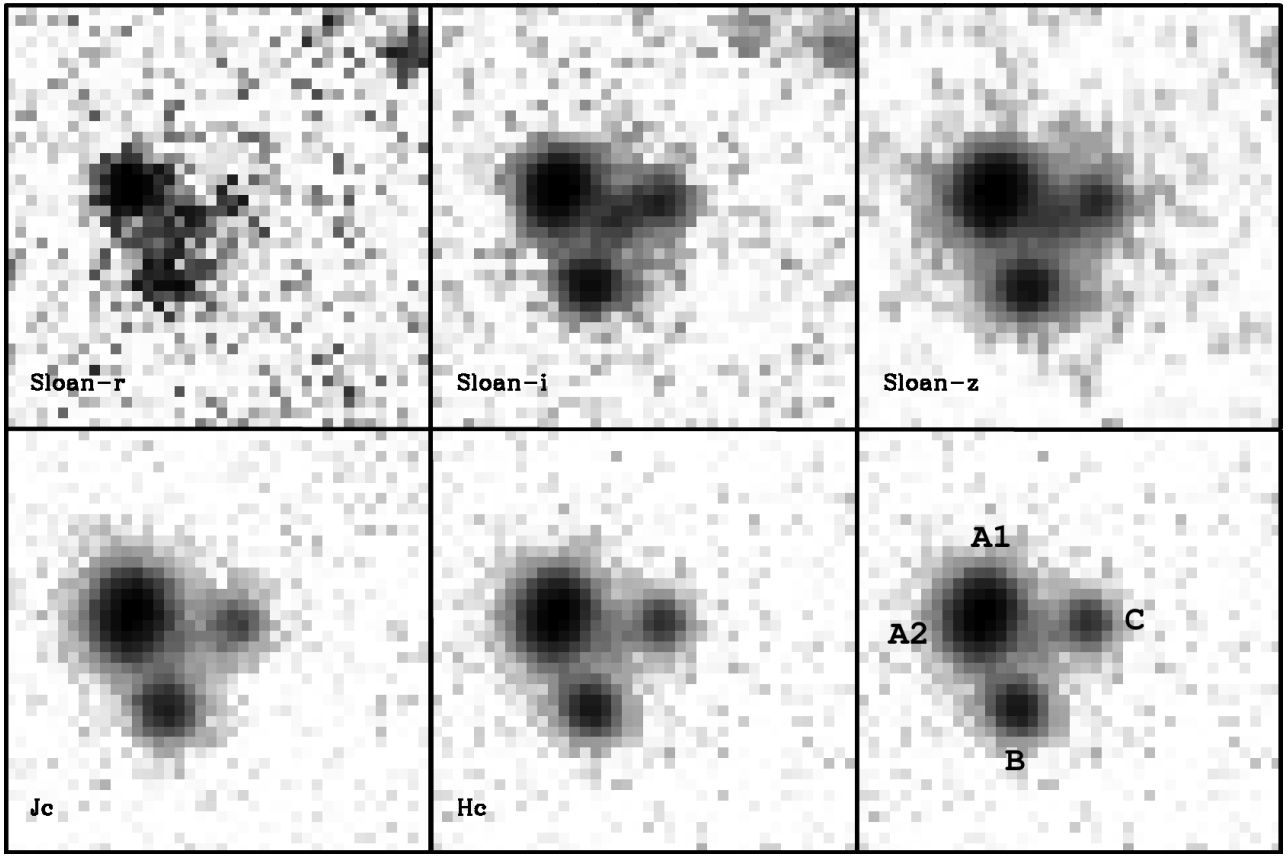}
  \caption{Magellan IMACS and PANIC imaging of MG 0414+0534, taken on 2007 November 3. The anomalous flux ratio $F_{obs}$ increases as we move blueward.}
  \label{images}
\end{figure*}

We observed MG 0414+0534 using the Magellan 6.5-metre Baade telescope on 2007 November 3, in $\la0\farcs5$ seeing. Observations were made using the IMACS (in f/2 mode) $r^\prime$, $i^\prime$, and $z^\prime$ filters and the PANIC $J$ and $H$ filters. Integration times were approximately five minutes per filter. Images are provided in Figure \ref{images}.

We fit Gaussians to each quasar image. We began by assuming a round, Gaussian PSF, and then allow the ellipticity to vary as a free parameter. The $H$-band image best shows the separation of the lensed images, and this was initially used to constrain the position of each component in each filter. In the final fit we freed up position as well, resulting in a 5-D fit (x, y, amplitude, FWHM, and ellipticity) for each component using a downhill gradient technique. 

The apparent magnitudes of the four lensed images and the lensing galaxy observed on 2007 November 3 are provided in Table \ref{0414mags}. Flux ratios in each filter between the two images of interest, $F_{obs} = A_2/A_1$, are provided in Table \ref{0414obs}. This table contains flux ratios from the observations presented in this paper, as well as two sets of \textit{HST} observations, obtained from the CASTLES Survey web page\footnotemark[1] and \citet{fls97}.

Table \ref{0414obs} shows that the flux ratio between images $A_2$ and $A_1$ at any single epoch grows larger as wavelength increases. We have assumed that this anomaly is solely due to microlensing. An alternative explanation is differential extinction due to dust in the lens. It has been argued previously that dust reddening is able to completely explain the anomalous flux ratios in MG 0414+0534 (\citealt{l95}; \citealt{mcl98}; \citealt*{ogm08}). However, \citet{tk99} showed that the lens galaxy is a passively evolving early-type galaxy, and thus unlikely to be dust dominated. Furthermore, several lensing systems displaying a flux ratio anomaly similar to MG 0414+0534 are known to exist. Microlensing models predict, as is observed, that in such systems it is always the saddle point ($A_2$) image that will be suppressed. Differential extinction due to dust would not be expected to favour one image over another.

\begin{table}
\caption{MG 0414+0534 magnitudes}
\label{0414mags}
\begin{tabular}{cccccc}
\hline
 & \multicolumn{5}{|c|}{Filter} \\
 source & $H$ & $J$ & $z^\prime$ & $i^\prime$ & $r^\prime$ \\
\hline
$A_1$ & 17.87 & 19.41 & 21.92 & 23.41 & 25.40 \\
$A_2$ & 17.44 & 18.85 & 20.76 & 21.97 & 23.72 \\
$B$ & 20.31 & 21.51 & 27.20 & 28.71 & 30.01 \\
$C$ & 21.12 & 22.84 & 28.37 & 29.51 & 30.87 \\
$L$ & 19.27 & 20.59 & 25.91 & 27.72 & 31.08 \\ 
\hline
\end{tabular}

\medskip
Apparent magnitudes of the lensed images ($A_1$, $A_2$, $B$, and $C$) and the lensing galaxy ($L$) in MG 0414+0534. Observations were taken in five filters with the IMACS and PANIC instruments on the Magellan 6.5-m Baade telescope, 2007 November 3.
\end{table}

\begin{table}
\caption{MG 0414+0534 flux ratios}
\label{0414obs}
\begin{tabular}{cccc}
\hline
Filter & Central $\lambda$ (\AA) & $F_{obs} = \frac{A_1}{A_2}$ & Date \\
\hline
$H$ & 16500 & $0.67\pm0.05$ & 2007 November 3 \\
$J$ & 12500 & $0.60\pm0.2$ & 2007 November 3\\
$z^\prime$ & 9134 & $0.34\pm0.1$ & 2007 November 3\\
$i^\prime$ & 7625 & $0.26\pm0.1$ & 2007 November 3\\
$r^\prime$ & 6231 & $0.21\pm0.1$ & 2007 November 3\\
F110W & 11250 & $0.64\pm0.04$ & 1997 August 14 \\
F205W & 20650 & $0.83\pm0.03$ & 1997 August 14 \\
F675W & 6714 & $0.40\pm0.01$ & 1994 November 8 \\
F814W & 7940 & $0.47\pm0.01$ & 1994 November 8 \\
\end{tabular}

\medskip
Central wavelengths and observed (anomalous) flux ratios $F_{obs}$ between images $A_2$ and $A_1$ in each of nine filters. The 2007 November 3 observations were taken with the IMACS and PANIC instruments on the Magellan 6.5-m Baade telescope. The 1997 August 14 observations were taken with the NICMOS instrument on \textit{HST} (obtained from the CASTLES Survey web page http://cfa-www.harvard.edu/castles/). The 1994 November 8 observations were taken with the WFPC2 instrument on \textit{HST} (\citealt{fls97}).
\end{table}

\begin{figure}
  \includegraphics[width=80mm]{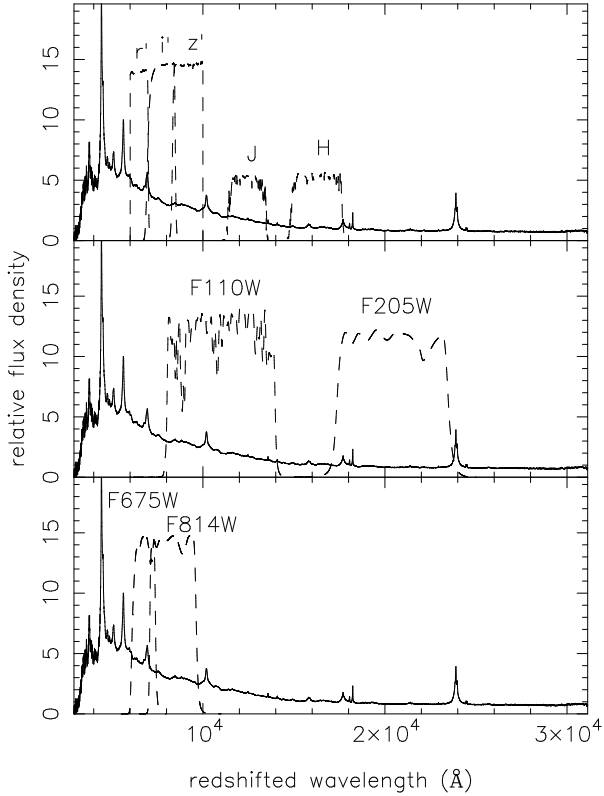}
  \caption{Composite spectrum constructed from 2204 quasar spectra selected from the Sloan Digital Sky Survey (\citealt{v01}). Overlaid (dashed line) are the transmission curves for the IMACS $r^\prime$, $i^\prime$, $z^\prime$ filters and the PANIC $J$ and $H$ filters (top panel), the \textit{HST} NICMOS F110W and F205W filters (middle panel), and the \textit{HST} WFPC2 F675W and F814W filters (bottom panel). The transmission curves have been rescaled by a constant factor for ease of comparison.}
  \label{filters}
\end{figure}

Broadband observations of quasar continuum emission are likely to be contaminated by broad emission lines (BELs), which may in turn affect a power-law fit to the radius of the accretion disc. As we expect the continuum and broad emission line regions to be differently microlensed, this contamination is difficult to quantify exactly. Using the SDSS composite quasar spectrum of \citet{v01}, we determined the BEL flux contribution to a (generic, unlensed) quasar at the same redshift as MG 0414+0534 in our wavebands to be less than 5 per cent. The composite spectrum is shown in Figure \ref{filters} with the filter transmission curves overlaid. The top panel shows the five Magellan IMACS and PANIC filters, the middle panel shows the two \textit{HST} NICMOS filters, and the bottom panel shows the two \textit{HST} WFPC2 filters used in our analysis.

Optical and IR spectra are available for MG 0414+0534 (\citealt{h92}; \citealt{a94}; \citealt{l95}). In these spectra, no BELs are detected within our wavebands. We therefore use the composite spectrum to illustrate the approximate size of BEL contamination. A detailed analysis of this source of error would require spectroscopy taken contemporaneously with each imaging epoch.
 
\section{Simulations}
We parameterise the frequency-dependent physical size of the accretion disc using Gaussian profiles of width $\sigma$ as a function of emitted wavelength $\lambda$:

\begin{equation}
\sigma = \sigma_0\left(\frac{\lambda}{\lambda_0}\right)^{\nu}
\label{powerlaw}
\end{equation}
The observational data provide us with flux ratio as a function of wavelength, $F(\lambda)$, at three separate epochs. Microlensing simulations, detailed below, offer the link between source size and flux ratio, $F(\sigma)$.

We conducted our microlensing simulations using an inverse ray-shooting technique (e.g. \citealt*{krs86}; \citealt*{wpk90a}). The key parameters for such simulations are the convergence $\kappa_{tot}$ and the shear $\gamma$ of the lens at the image positions. We used the MG 0414+0534 lens model of \citet{wms95}, provided in Table \ref{lens_para}. The convergence of the lens can be split into two components -- a compact stellar component $\kappa_*$ and a continuously distributed component $\kappa_c$. We allowed the smooth matter percentage $s = \kappa_c/\kappa_{tot}$ in the lens to vary between 0 per cent and 99 per cent, in 10 per cent increments. 

The microlenses were drawn from a Salpeter mass function $dN/dM \propto M^{-2.35}$ with $M_{max}/M_{min}=50$. Physical sizes are therefore scaled by the average Einstein Radius projected on to the source plane $\eta_0$, which is $3.75 \times 10^{16}h_{70}^{-1/2}(\langle M \rangle/\rmn{M}_{\odot})^{1/2}\rmn{cm}$ for MG 0414+0534. Magnification maps were generated covering an area of $24\eta_0 \times 24\eta_0$, with a resolution of $2048 \times 2048$ pixels. Twenty maps were generated for each image and for each smooth matter percentage.

\begin{table}
\caption{Lensing parameters}
\label{lens_para}
\begin{tabular}{lccl}
\hline
Image & $\kappa_{tot}$ & $\gamma$ & $\mu_{tot}$ \\
\hline
$A_1$ & 0.472 & 0.488 & 24.2 \\
$A_2$ & 0.485 & 0.550 & -26.8 \\
\hline
\end{tabular}

\medskip
Lensing parameters for images $A_1$ and $A_2$ in MG 0414+0534 (\citealt{wms95}).
\end{table}

We then randomly selected source positions in each of the $A_1$ and $A_2$ magnification maps. At each source position, we calculated the magnifications of 40 Gaussian sources of increasing characteristic width $\sigma$. The characteristic width of the sources varied from $0.05\eta_0$ to $2.00\eta_0$ in increments of $0.05\eta_0$. Dividing $A_1$ magnifications by $A_2$ magnifications allowed us to build up a library of 25500 curves of magnification ratio (equivalently, flux ratio) as a function of source width, $F_i(\sigma)$.

We allowed the width of the inner-most ($r^\prime$-band) emission region $\sigma_0$ to vary between $0.05\eta_0$ and $2.00\eta_0$, and the power-law index $\nu$ to vary between 0 and 15. These values were found to cover the interesting areas of parameter space in coarse simulations. For each combination of $\sigma_0$ and $\nu$, we calculated emission region widths $\sigma_n$ at the central wavelength of each filter, $\lambda_n$ (see Table \ref{0414obs}), using Equation \ref{powerlaw}. The value of $\lambda_0$ was taken to be 6231 \AA, the central wavelength of the $r^\prime$-filter.

By comparing $\sigma_n(\lambda_n)$ with each of our library of $F_i(\sigma)$ curves (interpolating linearly in $\sigma$), we obtained 25500 values of $F^{sim}_n(\lambda_n)$ for each combination of $\sigma_0$ and $\nu$ at each epoch. These were then compared with $F^{obs}_n(\lambda_n)$, to obtain $\chi^2_i$ values, which were converted to likelihoods using:

\begin{equation}
L_i(F^{obs}|\sigma_0,\nu, s) = \rmn{exp}\left(\frac{-\chi^2_i}{2}\right).
\label{chitolike}
\end{equation}
A final likelihood was obtained for each combination of $\sigma_0$ and $\nu$ at each epoch by summing over the individual likelihoods for each of the 25500 simulated flux ratio curves:

\begin{equation}
L(F^{obs}|\sigma_0,\nu, s) = \sum_{i=1}^{25500}L_i(F^{obs}|\sigma_0,\nu, s).
\label{finall}
\end{equation}

We convert these likelihoods to an a-posteriori differential probability distribution using Bayes' theorem: 

\begin{equation}
\frac{d^3P}{d\sigma_0d\nu ds} \propto L(F^{obs}|\sigma_0,\nu,s)\frac{dP_{prior}}{d\sigma_0} \frac{dP_{prior}}{d\nu}\frac{dP_{prior}}{ds}
\label{bayes}
\end{equation}
We assume constant Bayesian priors for inner width $\sigma_0$, power-law index $\nu$ and smooth matter percentage $s$. 

\begin{figure*}
  \includegraphics[width=170mm]{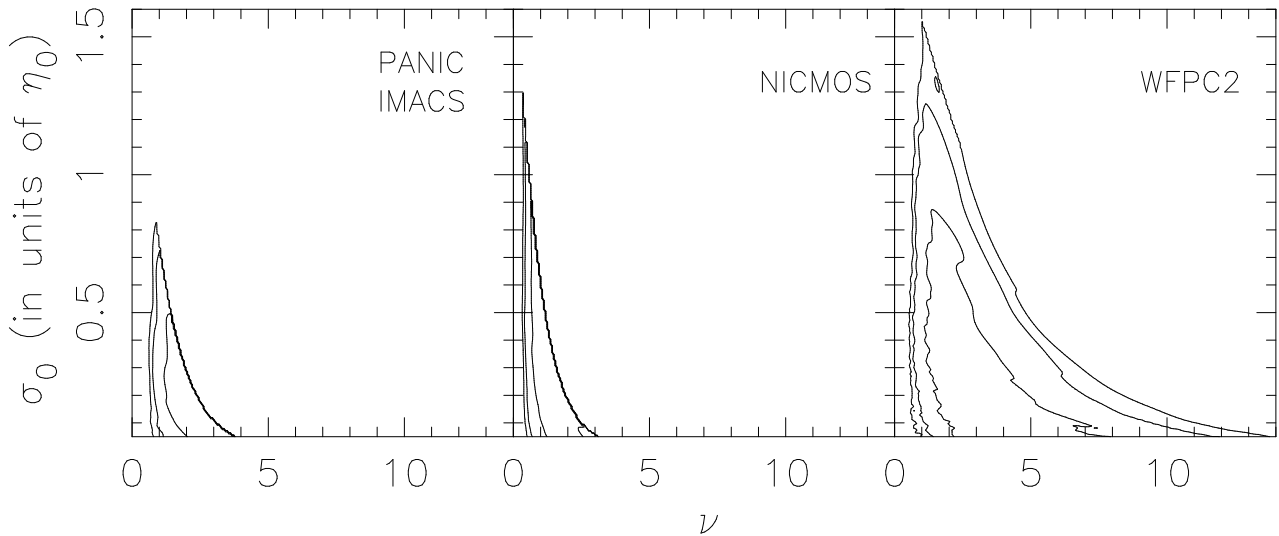}
  \caption{Probability distributions for width of the $r^\prime$-band emission region $\sigma_0$, in units of the average Einstein Radius $\eta_0$, and the power-law index $\nu$. Contours are 64, 26 and 14 per cent of peak probability. Distributions are shown for the Magellan IMACS and PANIC observations (left panel), the \textit{HST} NICMOS observations (middle panel), and the \textit{HST} WFPC2 observations (right panel).}
  \label{individual}
\end{figure*}

Marginalising over smooth matter percentage, we obtain probability distributions for $\sigma_0$ and $\nu$ at each epoch: 

\begin{equation}
\frac{d^2P}{d\sigma_0d\nu} = \int\frac{d^3P}{d\sigma_0d\nu ds} ds
\label{overk}
\end{equation}
The individual probability distributions are provided in Figure \ref{individual}. The left panel shows the Magellan IMACS and PANIC distribution, the middle panel shows the \textit{HST} NICMOS distribution, and the right panel shows the \textit{HST} WFPC2 distribution.

The probability distributions at each of the three epochs were then combined to obtain a final distribution for $\sigma_0$ and $\nu$, which is shown in Figure \ref{contours} (top panel).

\section{Results}

\begin{figure}
  \includegraphics[width=80mm]{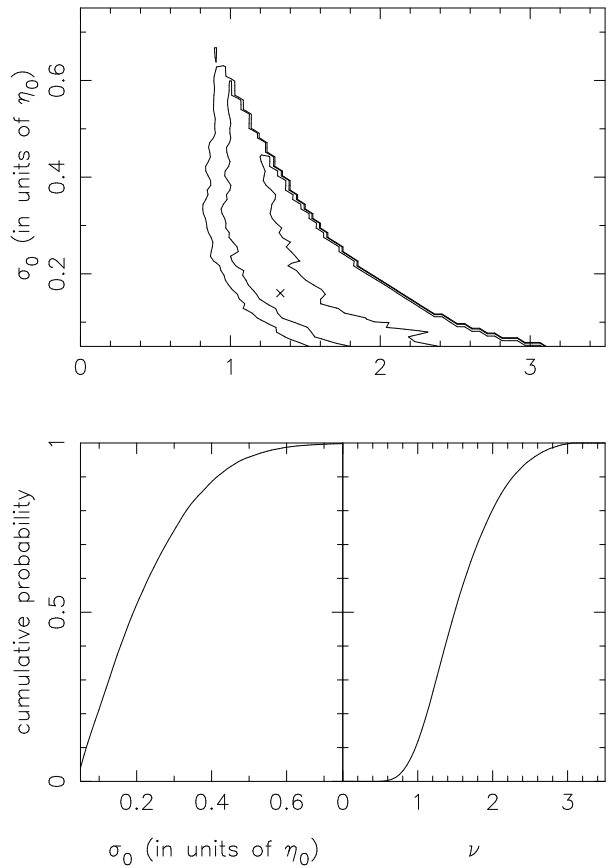}
  \caption{Top: probability distribution for width of the inner-most ($r^\prime$-band) emission region $\sigma_0$, in units of the Einstein Radius $\eta_0$, and the power-law index $\nu$. Contours are 64, 26 and 14 per cent of the peak probability. The cross marks the locus of a Shakura-Sunyaev disc with 20 per cent accretion efficiency and $M_{bh} = 1.82\times10^9\rmn{M}_{\odot}$ \citep{peng06}. Bottom left: cumulative probability distribution for the width of the $r^\prime$-band emission region $\sigma_0$, in units of the Einstein Radius $\eta_0$. Bottom right: cumulative probability distribution for the power-law index $\nu$.}
  \label{contours}
\end{figure}

Figure \ref{contours} (top panel) shows contours through the probability distribution for the Gaussian width of the inner-most ($r^\prime$-band) emission region $\sigma_0$ and power-law index $\nu$. By marginalising over the power-law index $\nu$ we obtain a 95 per cent upper limit on the width of the $r^\prime$-band emission region in MG 0414+0534 of $0.48\eta_0$ ($1.80 \times 10^{16} h_{70}^{-1/2}(\langle M\rangle/\rmn{M}_{\odot})^{1/2}$cm in physical units). Similarly, we can marginalise over $\sigma_0$ and obtain 95 per cent confidence limits on the power-law index of $0.77\leq\nu\leq2.67$. Marginalised cumulative probability distributions are provided in Figure \ref{contours} for $\sigma_0$ (bottom left panel) and $\nu$ (bottom right panel). The smooth matter percentage in this lens is unconstrained.

The standard Shakura-Sunyaev thin-disc model has $R \propto \lambda^{4/3}$ (ignoring the central temperature depression, which is located at scales smaller than smallest width probed in our analysis). This is consistent with our analysis at the 68 per cent confidence level.

Using a black hole mass for MG 0414+0534 of $M_{bh} = 1.83 \times 10^{9}\rmn{M}_{\odot}$, obtained using the virial technique with the $H\beta$ line \citep{peng06}, we can calculate the predicted half-light radius of the $r^\prime$-band emission region from a Shakura-Sunyaev thin disc model. We can compare this theoretical value with our observational results as microlensing fluctuations are sensitive only to the half-light radius of the emission region \citep*{msw05}. We assume a 20 per cent accretion efficiency, and an accretion disc inner radius equal to the innermost stable orbit. We find that the Shakura-Sunyaev thin disc for MG 0414+0534 in the $r^\prime$ band (1712\AA~in the quasar rest frame) has a half-light radius $R_{1/2} = 0.20 \eta_0$. This corresponds to a Gaussian source with characteristic width $\sigma_0 = 0.16\eta_0$. This size can be combined with $\nu = 4/3$, and is marked as a cross in Figure \ref{contours} (top panel).

There has been some disagreement between previous microlensing analyses regarding the size of optical emission regions in lensed quasars. \citet{pooley07} and \citet{metal08} both found that quasar optical emission regions are larger than predicted by Eddington limited thin disc models. Conversely, \citet{timo08} and \citet{poin08} found disc sizes consistent with thin disc models. 

\citet{poin08} noted that the solution to this discrepancy might be a slightly shallower temperature profile than predicted for a Shakura-Sunyaev thin disc. Our results are consistent with this suggestion. We note, however, that the only previous analysis of MG 0414+0534 \citep{pooley07} found that it was one of only two lensed quasars in a sample of ten for which a Shakura-Sunyaev disc was just large enough to explain both the observed X-ray and optical flux ratios.

\citet{poin08} also fit a power-law in wavelength to accretion disc radius for the lensed quasar HE 1104-1805. They obtained a power-law index of $\nu = 1.64^{+0.56}_{-0.46}$ (68 per cent confidence). At the 68 per cent confidence level, the power-law index obtained here is $\nu = 1.48^{+0.60}_{-0.43}$. These results are indistinguishable at the $1\sigma$ level. This is particularly interesting as the methods used in the two analyses differ significantly. \citet{poin08} compared 13 years of monitoring data in 11 wavebands to simulated microlensing light curves. We take advantage of the fact that two images straddling a critical curve each behave differently when microlensed, allowing us to obtain similar results with only 3 epochs of imaging in 10 overlapping wavebands.

\section{Conclusions}
In conclusion, we have demonstrated a technique for using near-contemporaneous multi-band imaging of anomalous lensed quasars to constrain the radius of the quasar continuum emission region as a function of emitted wavelength. The application of this technique to MG 0414+0534 yielded an upper limit on the width of the $r^\prime$-band emission region of $1.80 \times 10^{16} h_{70}^{-1/2}(\langle M\rangle/\rmn{M}_{\odot})^{1/2}$cm, and a power-law index range of $0.77\leq\nu\leq2.67$, both with 95 per cent confidence. These results are consistent with a Shakura-Sunyaev accretion disc model.

The greatest strength of our method is its observational efficiency. Previous analyses have relied upon multi-epoch monitoring, with the tightest size constraints typically obtained during the fortuitous observation of high magnification events. The method demonstrated here requires only as much time as is necessary to obtain sufficient signal to noise of the lensed images in each filter. However, it is only useful if the lensed quasar is displaying an anomalous flux ratio. We have assumed this anomalous flux ratio is produced solely by microlensing, although we cannot rule out differential extinction due to dust in the lensing galaxy as an alternative explanation.

We have obtained observations of additional systems with anomalous flux ratios, the analysis of which are forthcoming.

\section*{Acknowledgements}
We thank the referee for significant assistance in improving the presentation of the results in this paper. NFB acknowledges the support of an Australian Postgraduate Award. DJEF acknowledges the support of a Magellan Fellowship from Astronomy Australia Limited. We are indebted to Joachim Wambsganss for the use of his inverse ray-shooting code.

Based on observations with the NASA/ESA Hubble Space Telescope, obtained at the Space Telescope Science Institute, which is operated by the Association of Universities for Research in Astronomy, Inc. (AURA), under NASA contract NAS5-26555.

\appendix

\bsp

\label{lastpage}

\end{document}